\documentclass[preprint,showpacs,preprintnumbers,amsmath,amssymb]{revtex4}

\usepackage{graphicx}
\usepackage{dcolumn}
\usepackage{bm}

\begin{document}

\title{Current control in a tilted washboard potential via time-delayed feedback}

\author{D. Hennig}
\affiliation{Institut f\"{u}r Physik, Humboldt-Universit\"{a}t
 zu Berlin, Newtonstr.~15, 12489 Berlin, Germany}

\date{\today}

\begin{abstract}
\noindent We consider motion of an overdamped Brownian particle 
in a washboard potential exerted to a static tilting force. The bias yields directed net particle motion, i.e. a current.  
It is demonstrated that with an additional time-delayed
feedback term the particle current can be reversed against the  direction of the bias. 
The control function induces a ratchet-like effect that 
hinders further current reversals and thus the particle moves against the direction of the static bias. Furthermore, varying the delay time allows also  to continuously depreciate and even stop the transport in the washboard potential.  
We identify and characterize the underlying mechanism which applies to current control in a wide temperature range. 
\end{abstract}

\maketitle



The dynamics of particle motion in periodic potentials at finite temperature is an extensively studied field \cite{Risken}. Being exerted to a static tilt the system responds with a permanent transport in the direction of the bias. 
A number of experimental situations can be described on the basis of  
one-dimensional particle motion in a tilted spatially periodic potential \cite{Faucheux}-\cite{Regtmeier}.
Here we consider particle motion in a washboard potential which is often employed as as a paradigm to model transport in one-dimensional periodic and symmetric
structures \cite{Risken},\cite{Faucheux}-\cite{Speer}. 
Our aim is to demonstrate that the direction of the current can be controlled  with the application of a time-delayed feedback method.
Although the delayed feedback method was originally proposed by Pyragas \cite{Pyragas} to stabilize unstable states in deterministic systems meanwhile it has been facilitated in various other contexts
\cite{handbook}.
Recently for the  control of transport in Brownian motors a feedback
strategy has been successfully utilized for two ratchet systems
interacting through a unidirectional delay coupling \cite{delay}.
The effect of time-delayed feedback on the rectification of thermal motion of  Brownian particles  
has been studied in overdamped ratchet systems \cite{Cao}. 
 
We consider an overdamped Brownian particle moving in a tilted 
one-dimensional spatially-periodic and symmetric potential. The dynamics is governed by
the following Langevin equation expressed in
dimensionless form
\begin{eqnarray}
\dot{q}(t)&=&-U^{\prime}(q(t))+F+\xi(t)+f(t)\,.\label{eq:qdot}
\end{eqnarray}
The dot and prime denote differentiation with respect to time and coordinate respectively. The potential is given by 
\begin{equation}
U(q)=U(q+1)=-\cos(2\pi q)/(2\pi)\,.
\end{equation} 
$F$ is a tilting force serving for a static bias. 
The particle is subjected to a Gaussian
distributed thermal, white noise $\xi(t)$ of vanishing mean
$\langle\xi(t)\rangle=0$, obeying the well-known
fluctuation-dissipation relation $\langle\xi(t)
\xi(t^{\prime})\rangle=2 k_B T\delta(t-t^{\prime})$ with
$k_B$ and $T$ denoting the Boltzmann constant and temperature
respectively. The last term in (\ref{eq:qdot}) denotes a continuous
feedback term of the form
\begin{equation}
f(t)=K(1-\tanh\left[q(t-\tau)-q(t)\right])\label{eq:feedback}
\end{equation}
of strength $K$ and with delay time $\tau$. Note the restriction $0\le f(t) \le 2K$.
In what follows we set exemplarily the value of the tilting force to $F=-0.75$. Motion in the biased periodic potential is then favored to the left. 
The potential barrier height is modulated by the static tilt as  
$\Delta E(F)=\{\sqrt{1-F^2}-|F|[\pi/2-\arcsin(|F|)]\}/\pi\,$.
For motion to the left with negative velocity $v$ there remains a potential barrier of $\Delta E(F=-0.75)=0.032$ to be surmounted. This has to be compared with the barrier height for the unbiased potential, $\Delta E(F=0)=1/\pi\simeq 0.318$, which we hereafter denote by $\Delta E$.

Without feedback, $K=0$, the particle forced by the thermal noise escapes from a well of the potential into an adjacent one (due to the negative tilt preferably to the left) where it may dwell for a while before a further escape happens. Nevertheless the long-term behavior reveals a negative net motion.
The expression for the corresponding average dwell time, $t_{dwell}$,  exhibits the typical exponential dependence on the ratio between the barrier height and the thermal energy according to
$t_{dwell}\varpropto \exp({\Delta E}/{k_BT})$ \cite{Kramers}.

Interestingly, this situation of net motion in one exclusive direction may change completely imposing the Langevin
dynamics to the time-delayed feedback. Indeed we found parameter
constellations for which the direction of the net motion  which,  
initially governed by the negative tilt, is eventually reversed due to the impact of the time-delayed feedback. 
In the numerical simulation of the system (\ref{eq:qdot}) with applied time-delayed feedback term (\ref{eq:feedback}) we set $f(t)=0$ in the interval 
$t\in [0,\tau)$, that is the system is affected by $f(t)$ only for $t\geqq\tau$. 
The impact of time-delayed feedback of strength $K=0.8$ is studied hereafter.
It is illustrative to introduce the {\it effective tilt force} as 
\begin{equation}
 F_{eff}(t)=K\left(1+\tanh[q(t)-q(t-\tau)]\right)+F\,.
\end{equation}
For times $t<\tau$ the particle evolves in the tilted washboard potential without feedback and moves preferably to the left. If $\tau$ is not too small (see further below) the coordinate difference $q(t)-q(t-\tau)$ may assume in the course of time such a large negative value that the feedback term, as given in Eq.\,(\ref{eq:feedback}), is almost zero. As long as the particle continues sliding down the tilted washboard potential this situation cannot change. On the other hand, if the particle dwells for long enough time, $t > \tau$, 
in a potential well the coordinate difference gets not only constrained but may even become positive.  Therefore the contribution from the feedback force, $K(1+\tanh[q(t)-q(t-\tau)])$,  to the effective tilt force, $F_{eff}(t)$, exceeds sufficiently the remaining one coming from the static tilt force, $F<0$, and motion in the positive direction is entailed. Provided that $K>|F|$ holds and if the motion in the right direction is sustained for a long enough period during which the coordinate difference can grow to a larger positive value, the bias of the washboard potential attains the value $F_{eff}(t)=2K+F>0$. This saturation level is then very likely maintained. 

Quantitatively the net particle motion is assessed by 
the current. The latter is measured by the average particle velocity which in the long-time limit is defined as
\begin{equation}
v= \lim_{t\rightarrow \infty} \frac{\langle{x}(t)\rangle}{t} \,,
\end{equation}
with $\langle \cdot \rangle$ indicating the average over the thermal noise.

The Langevin equation were numerically integrated by means of a two-step stochastic Heun solver scheme.
Without feedback we found that the current obeys the relation $v \simeq -0.163 \times E_{thermal}$ for a static tilt $F=-0.75$ where 
the thermal energy is measured in units of the (unbiased) barrier height, viz.  $E_{thermal}=k_BT/\Delta E$.

\begin{figure}
\includegraphics[scale=0.25]{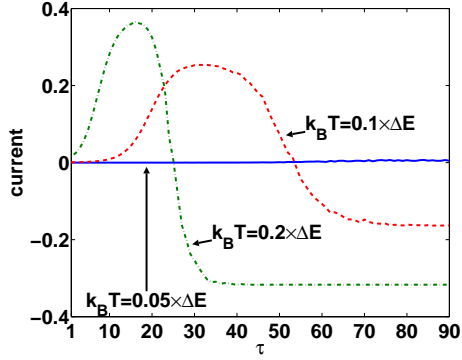}
\caption{\label{fig:Fig1} Current in
dependence of the delay time with fixed feedback strength $K=0.8$ and for different 
temperatures as indicated in the plot.
The static tilt force is given by $F=-0.75$.}
\end{figure}

\begin{figure}
\includegraphics[scale=0.25]{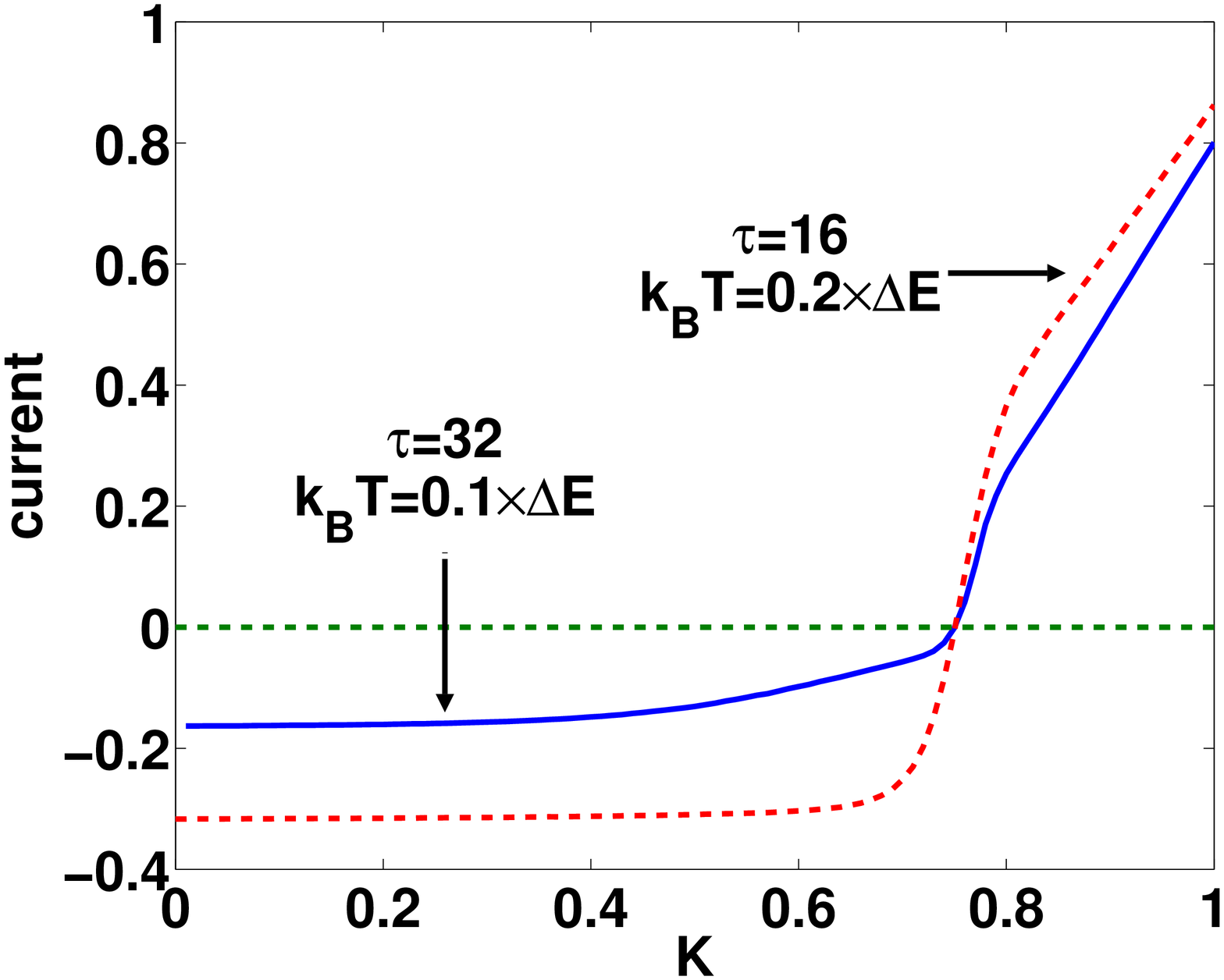}
\caption{\label{fig:Fig2} Current as a function of the feedback strength
$K$ for a fixed delay time and for different 
temperatures as indicated in the plot.
The static tilt force is given by $F=-0.75$.}
\end{figure}

\begin{figure}
\includegraphics[scale=0.25]{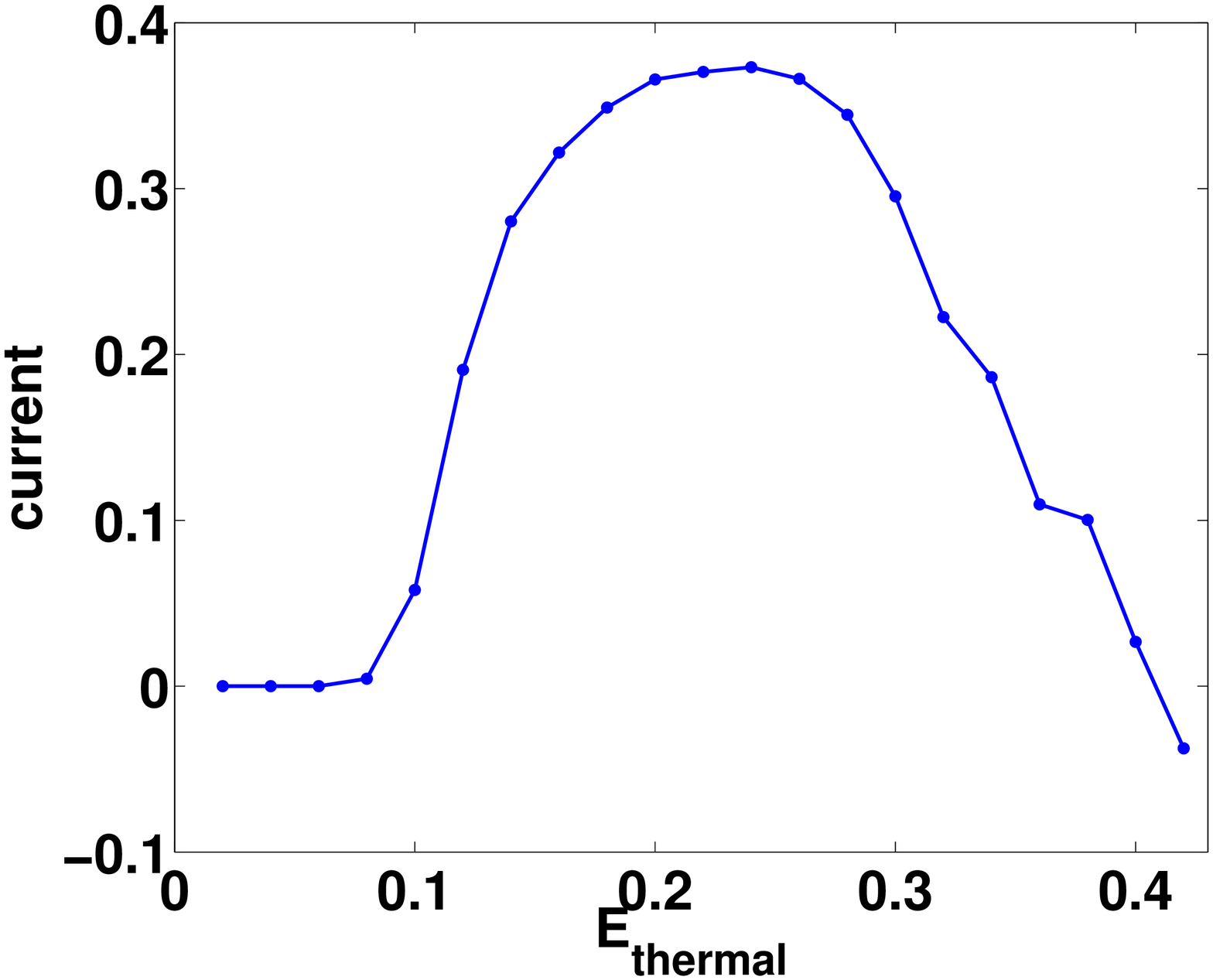}
\caption{\label{fig:Fig3} Current in dependence of the scaled thermal energy $E_{thermal}=k_BT/\Delta E$ for fixed feedback strength
$K=0.8$ and fixed delay time $\tau=16$. 
The static tilt force is given by $F=-0.75$.}
\end{figure}

To illustrate the features of the time-delayed feedback we show the average velocity as a function of the delay time $\tau$  for three different fixed temperatures in Fig.~\ref{fig:Fig1}. Averages were performed over $500$ realizations of the thermal noise.
For a low thermal energy $k_BT=0.05\times \Delta E$ the current virtually vanishes regardless of the delay time $\tau$ as long as $\tau \gtrsim 1$. Apparently, for delay times $\tau \gtrsim 1$ the feedback restrains the difference coordinate $q(t)-q(t-\tau)$ to small-amplitude fluctuations around zero. This in turn keeps the effective tilt force $F_{eff}=K+F$ small so that for low thermal energies barrier crossings are impeded. Thus with applied feedback term a transition from negative to zero current is achieved.
This changes considerably for higher temperatures where the current can be reversed due to applied time-delayed  feedback. More precisely for $k_BT\gtrsim 0.1\times \Delta E$ there exists an interval of delay times for which the current becomes positive. Moreover, as also seen in Fig.~\ref{fig:Fig1},  
within this interval  a resonance structure arises for which it holds that the higher the thermal energy the higher the {\it maximal} current and the more the position of the latter is shifted to a smaller $\tau_{peak}$ value. For each temperature there exist a critical value of the delay time beyond which the current becomes negative again.  Eventually for large enough $\tau$ values the  current attains the level adopted for zero feedback strength, $K=0$, that is the feedback is no longer of influence.
To understand further the effect of the feedback term we plot in Fig.~\ref{fig:Fig2} 
the current as a function of the feedback strength $K$ for two different values of the thermal energy.
For $k_BT=0.2\times \Delta E$ we notice that while for small but positive values of $K$ the current is not modified compared to its value adopted for $K=0$ there is a more or less sudden increase for $K \gtrsim 0.7$ and crossover from negative to positive values of $v$ takes place at $K=0.75$, that is when the two contributions to the effective tilt force $F_{eff}$ balance. Further enlargement of $K$ leads to a comparatively rapid linear growth of the current. The same holds true for the lower thermal energy $k_BT=0.1\times \Delta E$ except that the rapid growth of $v$ at the crossover point, which is also located at $K=0.75$, is not as pronounced as in the previous case. 
The current in dependence of the thermal energy $E_{thermal}=k_BT/\Delta E$,
measured in units of the barrier height, which is shown in Fig.~\ref{fig:Fig3} 
for an exemplary fixed feedback strength $K=0.8$ and delay time $\tau=16$,  exhibits a resonance structure. 
That is, there exist an {\it optimal} thermal energy, $E_{termal}\sim 0.24$, associated with a peak of the current  to either side of which the current decays fairly rapidly.  

\begin{figure}
\includegraphics[scale=0.2]{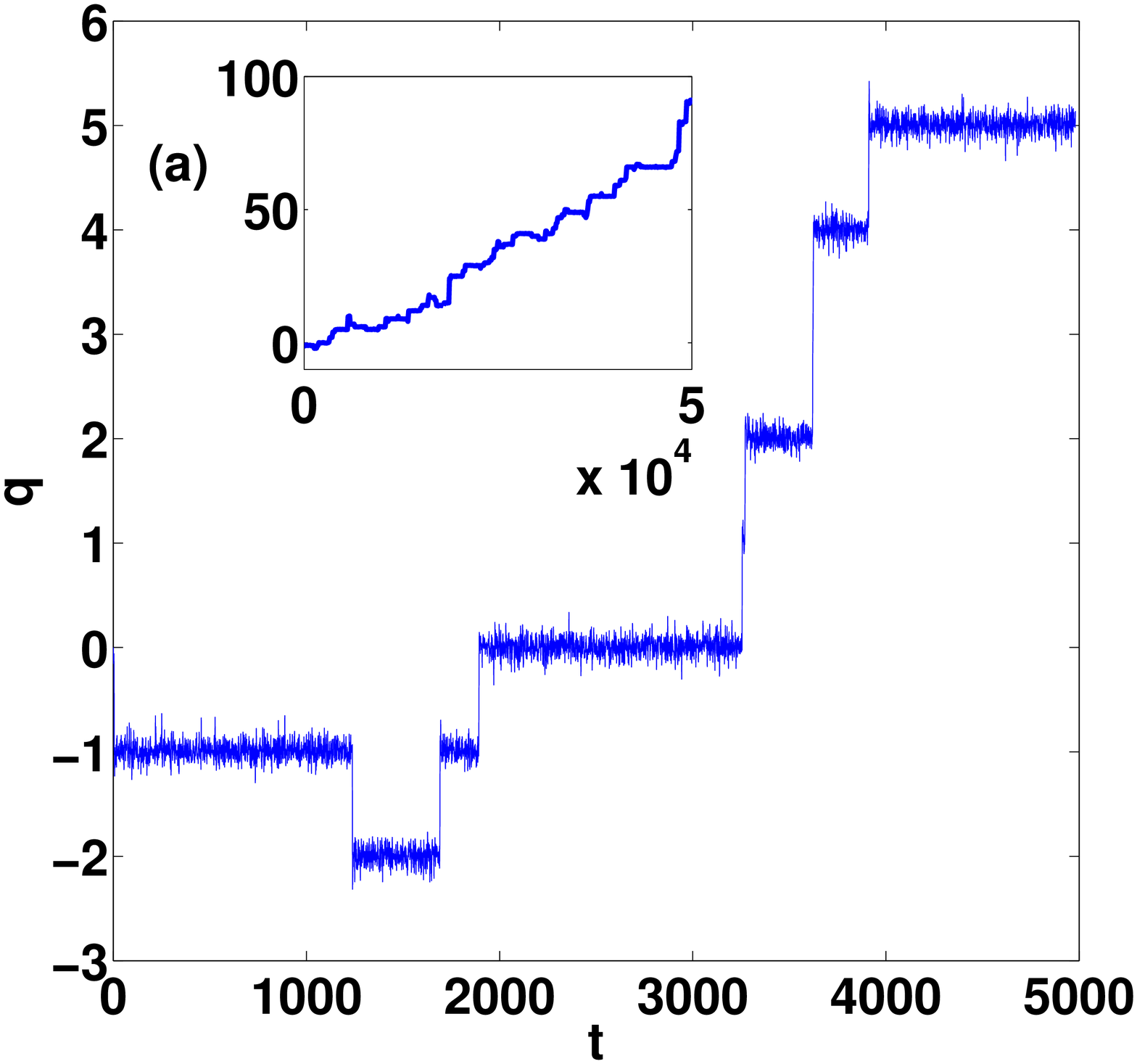}
\includegraphics[scale=0.2]{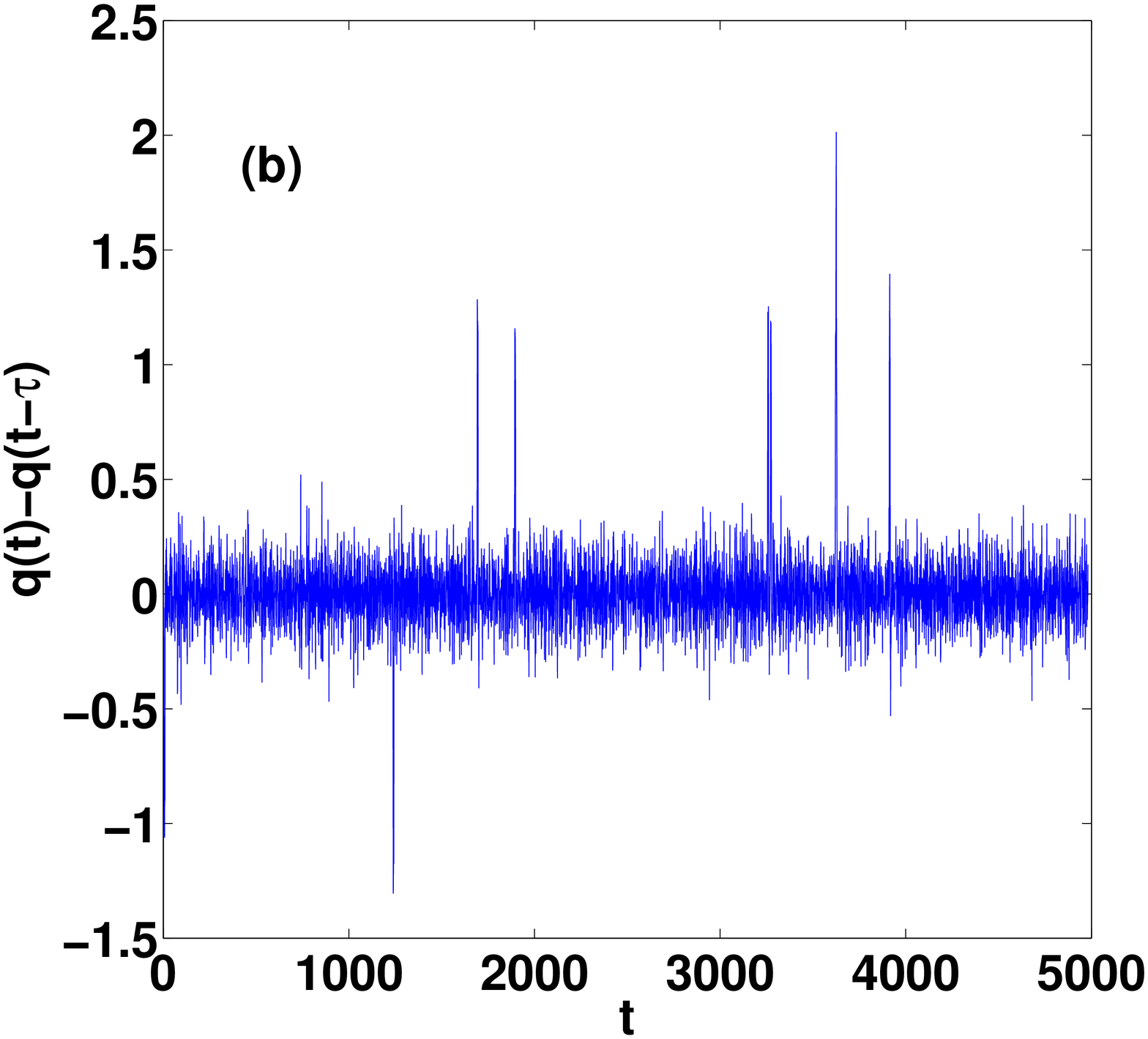}
\includegraphics[scale=0.2]{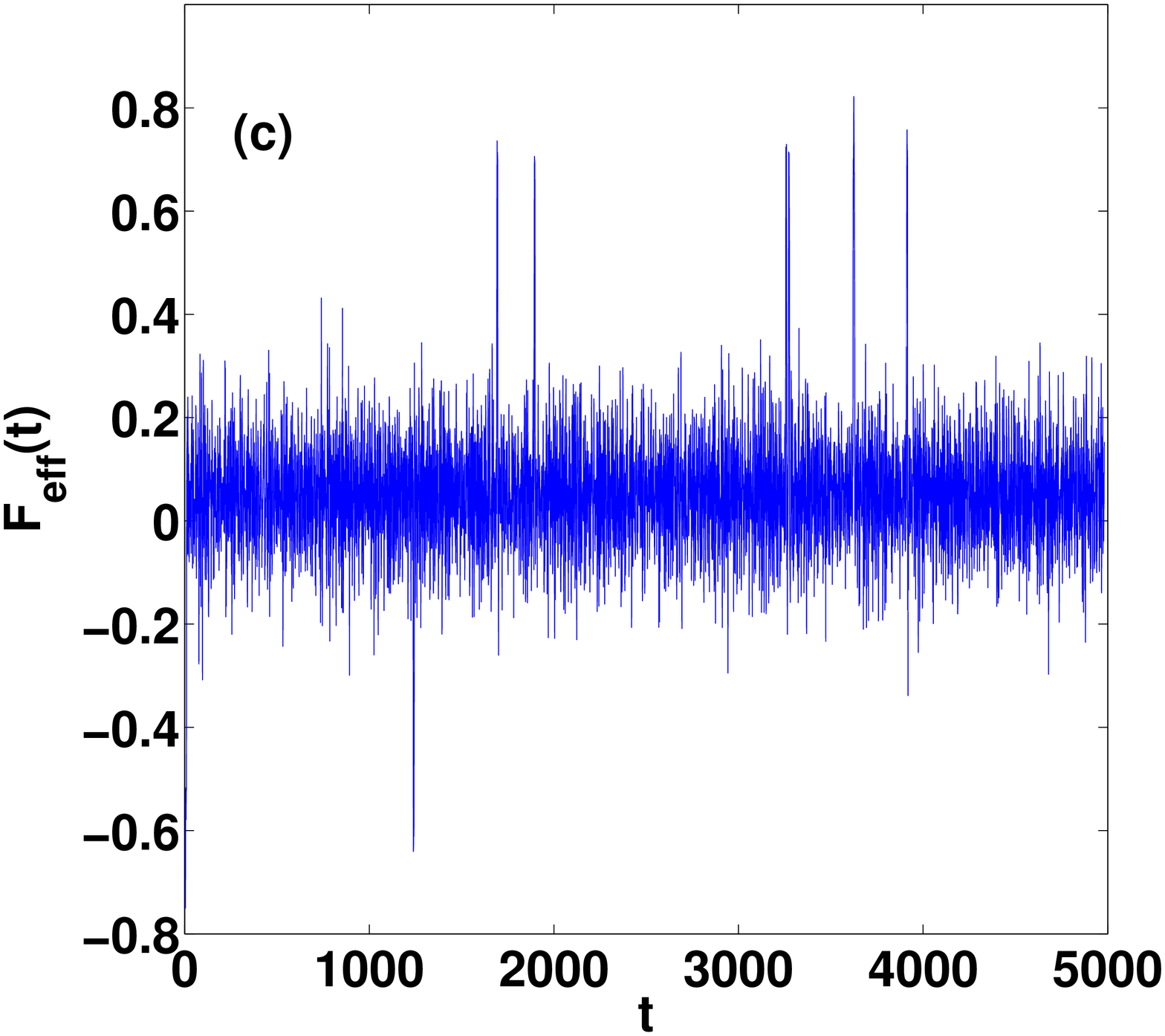}
\caption{\label{fig:Fig4} (a): Time evolution of the position for one realization of noise. The inset displays the long-term behavior.
(b): The difference $q(t)-q(t-\tau)$. (c): The effective tilt force $F_{eff}(t)$.
 Notice the assignment of the bursts in (b) and (c) to the respective particle transitions.
The parameter values read as: 
$F=-0.75$, $K=0.8$, $\tau=5$ and $k_BT=0.1\times \Delta E$.}
\end{figure}

\begin{figure}
\includegraphics[scale=0.2]{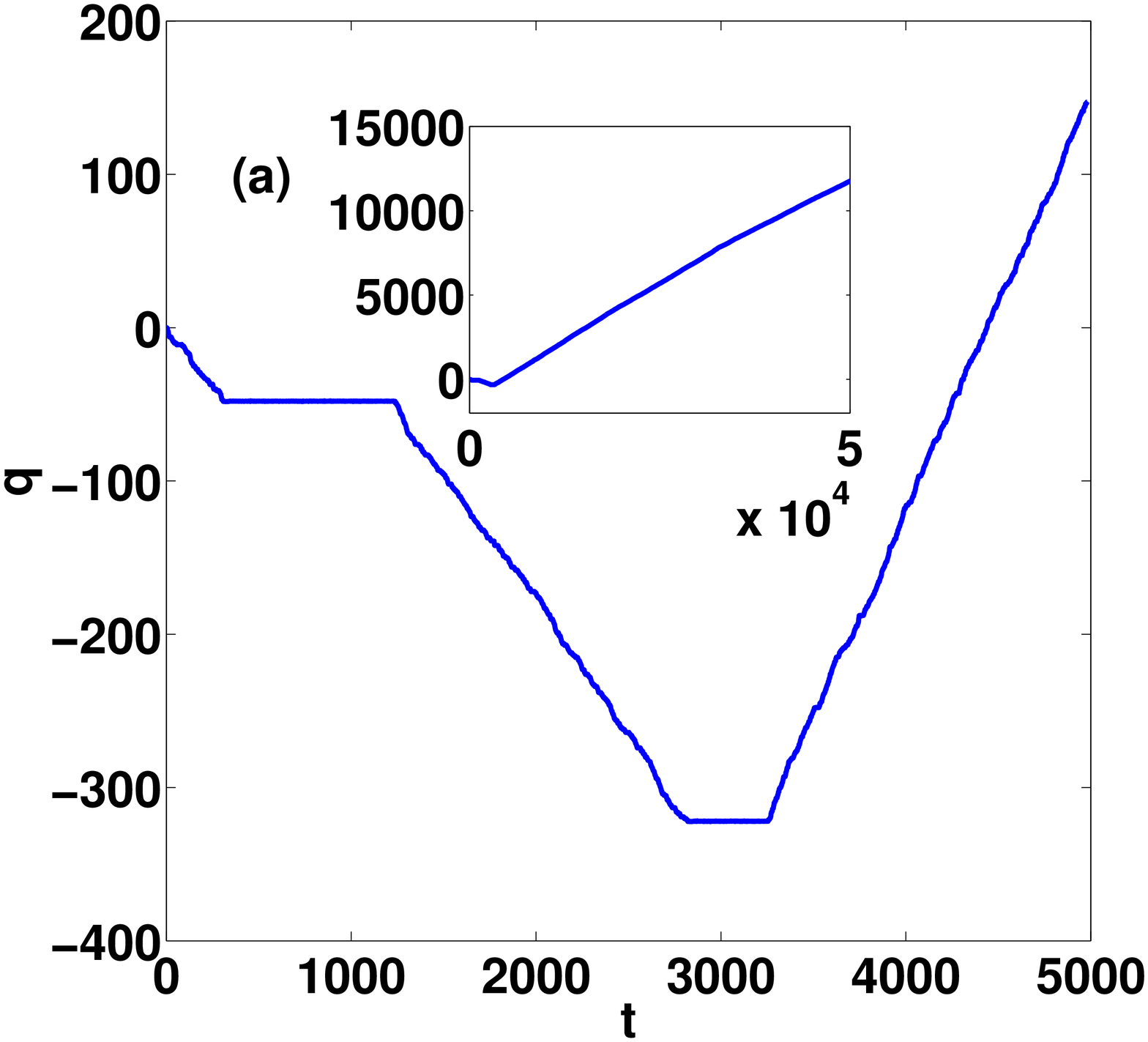}
\includegraphics[scale=0.2]{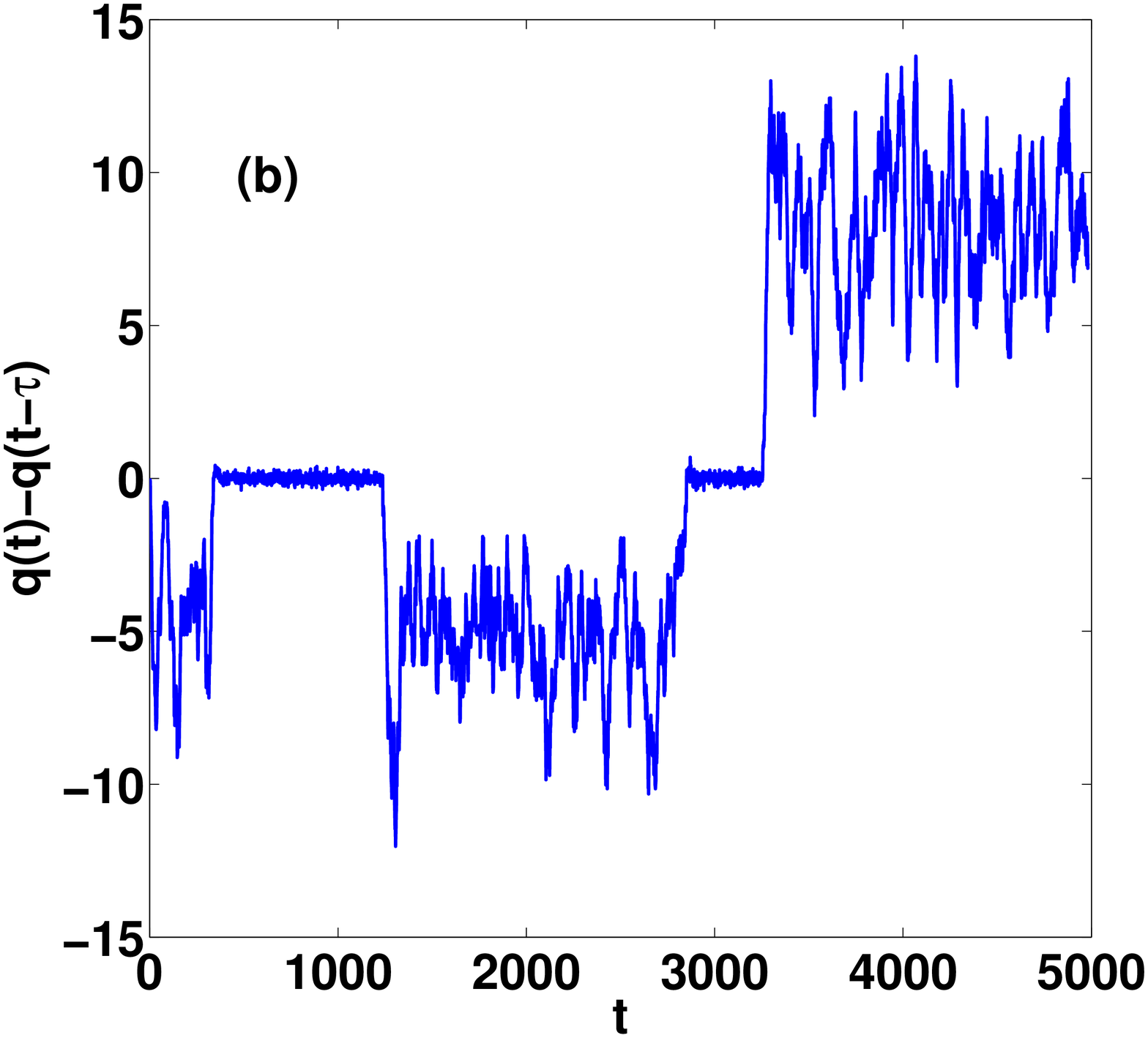}
\includegraphics[scale=0.2]{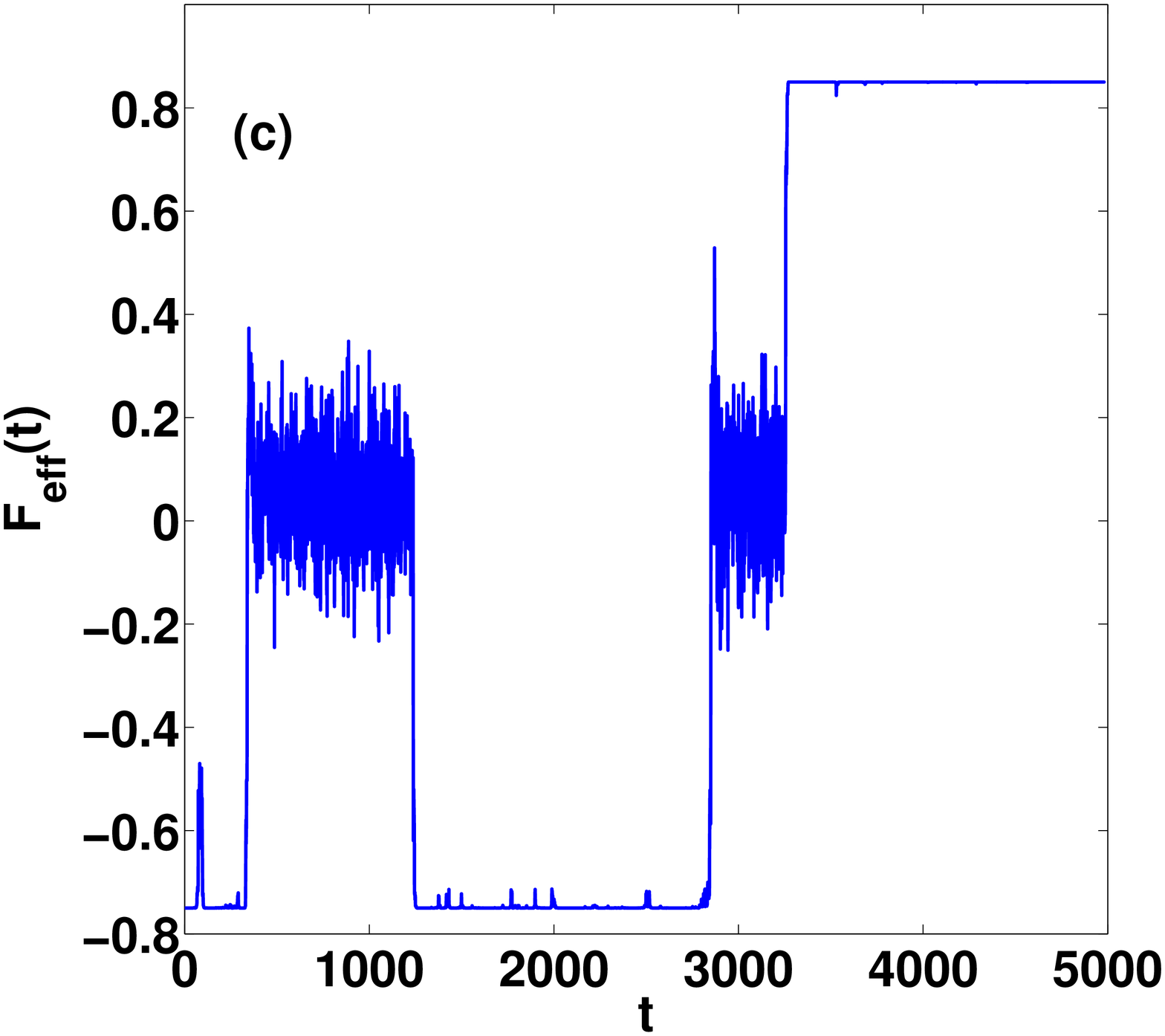}
 \caption{\label{fig:Fig5} 
As in Fig.~\ref{fig:Fig4} for the same realization of thermal noise but for  delay time $\tau=32$.
}
\end{figure}

To gain insight into the role played by the delay time  with regard to the current efficiency we represent the time evolution of the coordinate $q(t)$, the difference $q(t)-q(t-\tau)$, and the effective tilt force $F_{eff}(t)$ for a delay time $\tau=5$ and $\tau=32$ in Figs.~\ref{fig:Fig4} and \ref{fig:Fig5} respectively for one realization of thermal noise.
For a delay time $\tau\simeq t_{dwell}\simeq 5<\tau_{peak}=32$, which is of the order of the dwell time, the resulting current is comparatively low but positive. This is reflected in the temporal evolution of the coordinate in Fig.~\ref{fig:Fig5}\,(a). 
In an early stage of the dynamics (prior to the application of the feedback) the particle escapes from its starting position (the bottom of a potential well) into the well adjacent to the left where it dwells for considerably long time. After another jump to the left has taken place 
further occasional jumps follow in the opposite direction. That behavior is related with the time evolution of the effective tilt force $F_{eff}(t)$ as well as the difference coordinate $q(t)-q(t-\tau)$ which most of the time perform small amplitude fluctuations around zero apart from the occasional sudden bursts being correlated with the kicking of the particle from one potential well into another one. Nevertheless on average the particle motion proceeds to the right (cf. the long-term behavior of $q(t)$ in the inset). 

In comparison for the delay time $\tau_{peak}=32>t_{dwell}$, yielding the peak in the current (cf. Fig.~\ref{fig:Fig1}), the particle slides already far down the tilted potential in an early stage for $t\lesssim 328$.  During this time the coordinate difference $q(t)-q(t-\tau)$ stays strictly negative as seen in Fig.~\ref{fig:Fig5}. Hence the delay term itself, $K(t)=K\left(1+\tanh[q(t)-q(t-\tau)]\right)$, is vanishingly small and consequently the particle feels to the full extent the bare tilt force $F<0$. 
Subsequently to a trapping interlude a longer excursion in the negative direction occurs terminated by yet 
another trapping period. For the particle dwelling inside the potential well and being buffeted by the thermal noise, likewise in the previous case, $F_{eff}(t)$ as well as $q(t)-q(t-\tau)$ fluctuate around zero. Strikingly, in contrast to the previous case of short delay time $\tau=5$, in the present case of $\tau=32$ the coordinate difference retains not only a strictly positive but a by far larger value for a comparatively longer time which in turn serves for ongoing positive effective tilt force $F_{eff}(t)\simeq 2K+F>0$. 
Conclusively, it is the interplay between the coordinate difference with persisting positive amplitude and the positive effective force that is responsible for forcing vigorously the particle in the right direction. 
Note that after current reversion the washboard potential experiences effectively a positive tilt, $2K+F>0$, that is larger than the absolute value, $|F|$, of the negative tilt started with. On the other hand, an increased tilt is related with a shorter average dwell time $t_{dwell}$ which makes it less likely that the coordinate difference returns to small values let alone become again negative. In this sense the asymmetry $K>|F|$ serves for a ratchet-like effect hampering a further reversal of the current back from positive to negative value.  

In conclusion, we have identified and characterized a mechanism to control the thermally driven transport for overdamped Brownian particles evolving in a  
washboard potential under the mutual impact of static bias and time-delayed feedback. 
That is with appropriately chosen strength of the delay term and the delay time the direction of the current can be reversed in such a way that particle motion proceeds opposite to the starting direction governed by the static bias. 
Due to its construction the control function induces a ratchet-like effect that 
hinders further current reversals. Moreover, varying the delay time allows also  to continuously tune the current between its possible negative and positive
extreme value including stoppage of transport in the washboard potential.  
Our results apply to a wide temperature range.

\end{document}